\documentclass[preprint,aps,floats,nofootinbib]{revtex4}

\input epsf.tex
\def\DESepsf(#1 width #2){\epsfxsize=#2 \epsfbox{#1}}

\usepackage{graphicx}
\usepackage{amsmath}
\usepackage{psfig}
\def\bea{\begin{eqnarray}}
\def\eea{\end{eqnarray}}

\def\thebibliography#1{\centerline{\bf REFERENCES}
  \list{[\arabic{enumi}]}{\settowidth\labelwidth{[#1]}\leftmargin
  \labelwidth\advance\leftmargin\labelsep\usecounter{enumi}}
\def\newblock{\hskip .11em plus .33em minus -.07em}\sloppy
  \clubpenalty4000\widowpenalty4000\sfcode`\.=1000\relax}

\begin{document}

\draft

\vspace*{0.5cm}

\title{$B_s^0 - \overline{B}_s^0$ Mixing in Leptophobic $Z^\prime$ Model}

\author{
Seungwon~Baek$^1$\footnote{sbaek@cskim.yonsei.ac.kr}, ~
Jong~Hun~Jeon$^1$\footnote{jhjeon@cskim.yonsei.ac.kr}, ~
and C.~S.~Kim$^{1,2}$\footnote{cskim@yonsei.ac.kr}}

\affiliation{
 {\rm 1:} Department of Physics, Yonsei University,
Seoul 120-479, Korea\\
 {\rm 2:} IPPP, Department of Physics, University of Durham, Durham DH1 3LE, England
 ~\vspace{1cm} }
\date{\today}

\begin{abstract}
\noindent
Leptophobic $Z^\prime$ gauge boson appears naturally in many grand
unified theories, such as flipped SU(5) or string-inspired
$E_6$ models. This elusive particle easily escapes the
direct/indirect detections because it does not couple to charged
leptons. However, it can generate flavor changing neutral current
at tree level. In this letter, we show that the
recently measured mass difference, $\Delta m_s$, in the
$B_s^0 -\overline{B}_s^0$ system improves the previous bound of flavor changing
effective coupling by about one order of magnitude, $i.e$ irrespective of its phase,
$|U_{sb}^{Z^\prime}| \leq 0.036$ for  $M_{Z^\prime} = 700$ GeV, and
$|U_{sb}^{Z^\prime}| \leq 0.051$ for  $M_{Z^\prime} = 1$ TeV.
\end{abstract}
\maketitle


\section{introduction}

In the standard model (SM), the flavor changing neutral current
(FCNC) processes first occur at one-loop diagrams.
Its rate is suppressed by small electroweak gauge coupling, CKM matrix
elements and loop factors. Therefore, these rare processes are very
sensitive probe of new physics (NP) beyond the SM because
some of these suppression factors can be lifted in general NP models.

Asymmetric $B$--factories and Tevatron have produced lots of $B$-mesons
and some rare
$B$-decays induced by FCNC have been measured with enough precision to
probe NP models.
Among them, the processes with $b \to s$ transition at
quark level, such as $B \to \pi K$~\cite{B2piK},
$B \to \rho (\phi) K^*$~\cite{B2roK},
$B \to \phi K_S$~\cite{B2phiK}, $B_s \to \mu^+ \mu^-$~\cite{Bs2mumu},
have attracted much interest because
they still allow much room for large NP contributions and
some of them show possible deviations from
the SM predictions.

Recently D{\O}~\cite{D0} and CDF~\cite{CDF} collaborations at
Fermilab Tevatron reported the first observation of another $b \to s$ FCNC
process, {\it i.e}, the mass difference $\Delta m_s$ in the $B_s^0 -
\overline{B}_s^0$ system:
 \begin{eqnarray}
  \text{D{\O}}~&:&~\quad 17 ~ \text{ps}^{-1} < \Delta m_s < ~21 ~ \text{ps}^{-1}
                        ~~\left( 90 \% ~ \text{C.L.}\right) ,
  \nonumber\\
  \text{CDF}  ~&:&~\quad \Delta m_s = 17.33_{-0.21}^{+0.42} (\text{stat.})
                                    \pm 0.07 (\text{syst.})~ \text{ps}^{-1}.
 \end{eqnarray}
Although these measurements are a little bit smaller than the SM
expectations, considering large hadronic uncertainties we cannot
strongly argue that it is a NP signal at the moment. They,
however, may give strong constraints on the NP models, which
predict $b \to s$ FCNC transitions. After the release of these new
experimental results, their implications have been considered in
many papers both in model independent
approach~\cite{model_indep1} and in specific NP
models including $Z^\prime$-model~\cite{zprime}, minimal
supersymmetric standard model (MSSM)~\cite{MSSM}
and warped extra dimension model~\cite{RS-kim}.

In this letter, we consider the implications of $\Delta m_s$ measurements
on leptophobic $Z^\prime$-model.
Leptophobic $Z^\prime$ gauge bosons appear naturally in many grand
unified theories (GUTs), {\it e.g.} flipped SU(5) or string-inspired $E_6$ GUT models.
In some scenarios FCNC occurs at tree-level $Z$ and/or $Z'$ couplings.

First, we briefly introduce the leptophobic $Z^\prime$ models
which lead to tree level FCNCs in Sec.~\ref{section2}. In
Sec.~\ref{section3} we present the relevant formulas for the
$B_s^0 - \overline{B}_s^0$ mixing and perform numerical analysis.
Concluding remarks are given also in Sec.~\ref{section3}.

\section{Leptophobic $Z^\prime$ model and FCNC}
\label{section2}

Leptophobic $Z^\prime$ gauge boson (leptophobia) occurs naturally
in flipped SU(5)$\times$U(1) scenario~\cite{Lopez:1996ta}. In
this model the spinor ({\bf 16}) representation of SO(10)
is decomposed under SU(5)$\times$U(1) as
\bea
 {\bf 16} \to ({\bf 10},1) + ({\bf \bar{5}},-3) + ({\bf 1},5).
\eea
The SM particles are contained in
\bea
 {\bf 10} = \{Q, d^c, \nu^c \}, \quad {\bf \bar{5}} = \{L, u^c\}, \quad {\bf 1}=\{e^c\}.
\eea
The $Z^\prime$ becomes leptophobic if ${\bf \bar{5}}$ and ${\bf 1}$ are
uncharged under the new U(1)$^\prime$. It is noted that the U(1)$^\prime$
charges of representation {\bf 10} can be generation-dependent in string
models. This induces $Z^\prime$-mediated FCNCs at tree level in the down-type quark
sector and/or left-handed up-type quark sector.

Another scenario for leptophobia is $E_6$ model with kinetic mixing.
In GUT or string-inspired point of view, the $E_6$ model is a very
plausible extension of the SM \cite{Rizzo:1998ut}. It is natural
that  a U(1)$^\prime$ gauge group remains as a low energy
effective theory after the symmetry breaking of the $E_6$ group.
We assume that the $E_6$ group is broken through the following
breaking chain
\begin{eqnarray}
&&E_6 \to {\rm SO(10)} \times {\rm U(1)}_\psi \nonumber \\
&&\phantom{E_6 }  \to {\rm SU(5)} \times {\rm U(1)}_\chi \times {\rm U(1)}_\psi \nonumber \\
&&\phantom{E_6 }  \to {\rm SU(2)}_L \times {\rm U(1)} \times {\rm U(1)}^\prime ,
\end{eqnarray}
where U(1)$^\prime$ is a linear combination of two additional
$U(1)$ gauge groups with
$$
Q^\prime = Q_\psi \cos \theta - Q_\chi \sin \theta,
$$
where $\theta $ is the familiar $E_6$ mixing angle.

The most general Lagrangian, which is  invariant under the SM gauge group
with an extra U(1)$^\prime$,
allows the kinetic mixing term
$\displaystyle {\cal L}_{\rm mixing} = - \frac{\sin \chi}{2} \tilde{B}_{\mu\nu}
\tilde{Z}^{\prime ~\mu\nu} $ between the U(1) and U(1)$^\prime$ gauge boson fields.
This off-diagonal term can be removed by the non-unitarity transformation
\begin{equation}
\tilde{B}_\mu = B_\mu - \tan \chi Z_\mu^\prime,
~~\tilde{Z}_\mu^\prime = \frac{Z_\mu^\prime}{\cos \chi}~,
\end{equation}
which leads to the possibility of leptophobia of the physical $Z^\prime$ gauge boson
with the $E_6$ mixing.
Once all the couplings are GUT normalized,
the interaction Lagrangian of fermion fields and $Z^\prime$ gauge
boson can be written as
\begin{equation}
{\cal L}_{\rm int} = - \lambda \frac{g_2}{\cos \theta_W}
\sqrt{\frac{5 \sin^2 \theta_W}{3}}
\bar{\psi} \gamma^\mu \left( Q^\prime + \sqrt{\frac{3}{5}}\delta Y_{SM} \right)
\psi Z_\mu^\prime ~,
\end{equation}
where the ratio of gauge couplings $\lambda = g_{Q^\prime}/g_Y$,
and $\delta=-\tan \chi/\lambda$ \cite{Rizzo:1998ut}.
The general fermion-$Z^\prime$ couplings depend on two free parameters,
$\tan \theta$ and $\delta$, effectively \cite{Babu:1996vt}.
The $Z^\prime$ boson can be
leptophobic when $(Q^\prime + \sqrt{\frac{3}{5}}\delta Y_{SM})=0$
for $L$ and $e^c$ simultaneously.
There are several ways to embed the SM particles and exotic fermions to the
fundamental {\bf 27} representation of $E_6$ \cite{Rizzo:1998ut,Leroux:2001fx}
while keeping the $Z^\prime$ leptophobic.

As mentioned above, the $Z(Z^\prime)$ couplings to quarks can
generate tree-level FCNCs. In general there can be four different
types of FCNCs in the down-type quarks as $Z(Z^\prime)$ couples to
left-(right-)handed down-type quarks. Since the $Z$-mediating FCNC
is too dangerous, we suppress them. This can be achieved as
follows; In $E_6$ model, the exotic fermion $h^c$ has the same
U(1)$_Y$ charges with $d^c, s^c, b^c$, and $Z$-mediating FCNCs in
the right-handed down-type quarks are absent. To suppress
$Z$-mediating FCNCs in the left-handed down-type quarks we assume
the unitary matrix $V_L^d$ diagonalizing the down-type mass matrix
is an identity matrix. In the flipped SU(5) model, if we assume
there is no $Z-Z^\prime$ mixing, the $Z$-mediating FCNC disappears
as in the SM.

Now we turn to $Z^\prime$-mediating FCNCs: The assumption,
 $V_L^d = 1$, in the $E_6$ model
automatically suppresses the FCNCs in the left-handed down-type quarks.
In flipped SU(5) model we can adopt the same assumption
and suppress the FCNCs in the same sector.
Then only $Z^\prime$-mediating FCNCs in the right-handed down-type quarks
survive.
After integrating out degrees of freedom of heavy exotic fermions
and gauge bosons,
the FCNC Lagrangian for the $b\to q (q=s,d)$ transition can be parameterized as
\begin{equation}
{\cal L}_{\rm FCNC}^{Z^\prime} = - \frac{g_2}{2 \cos \theta_W}
U_{qb}^{Z^\prime} \bar{q}_R \gamma^\mu b_R Z_\mu^\prime ,
\end{equation}
where all the theoretical uncertainties including the mixing parameters
are absorbed into the coupling $U_{qb}^{Z^\prime}$.

%
%

The constraints on the $U_{qb}^{Z^\prime}$ were previously
considered in \cite{Leroux:2001fx} and \cite{Jeon:2006nq}.
In \cite{Jeon:2006nq} it was demonstrated that
the exclusive semi-leptonic $B \to  M\nu_R\bar{\nu}_R$  decays
give similar but stronger bounds than those obtained in \cite{Leroux:2001fx}.
It should be noted that the leptophobic $Z^\prime$ scenario is not
constrained at all by (semi-)leptonic decays, $b \to s \ell^+ \ell^-$
or $B_{(s)} \to \ell^+ \ell^-$
which strongly constrains typical NP models such as MSSM.
It is simply because leptophobic $Z^\prime$ does not couple to
ordinary leptons. This feature also distinguishes leptophobic
$Z^\prime$ models from other $Z^\prime$ models \cite{zprime}.

\section{$B_s^0 -\overline{B}_s^0$ Mixing and Constraints on $U_{qb}^{Z^\prime}$}
\label{section3}

Within the SM, the mass difference between the mass eigenstates in the
$B_s^0 -\overline{B}_s^0$ system is
\begin{eqnarray}
\Delta m_s^{\rm SM}
 = \frac{G_F^2}{6 \pi^2}
        M_W^2 m_{B_s}
        \eta_B S_0(x_t)
        \left( f_{B_s} \hat{B}_{B_s}^{1/2} \right)^2
        \left( V_{tb} V{_{ts}^\ast} \right)^2,
\end{eqnarray}
where $G_F$ is the Fermi constant, $\eta_B$ is a short-distance QCD correction,
$f_{B_s}$ is the decay constant for $B_s$ system, and $\hat{B}_{B_S}$ is the bag
parameter defined as
\begin{eqnarray}
\hat{B}_{B_s}
 = B_{B_s} (\mu_b)
   \left[ \alpha_s^{(5)}(\mu_b) \right]^{-6/23}
   \left[ 1 +\frac{\alpha_s^{(5)}(\mu_b)}{4 \pi} J_5 \right].
\end{eqnarray}
Main theoretical uncertainties arise from the hadronic parameter
$f_{B_s} \hat{B}_{B_s}^{1/2}$.


Using the CKMfitter results presented at FPCP06
\cite{Charles:2004jd}, $|V_{tb}| \simeq 1 , ~~ |V_{ts}| =
0.04113_{-0.00062}^{+0.00063}$, $M_{B_s}$ = 5.3696 GeV, $\eta_B
\simeq 0.551$, $S_0(x_t)=2.29_{-0.04}^{+0.05}$ with $\overline{m}_t (m_t)
= 162.3 \pm 2.2$ GeV and the hadronic parameter
\begin{eqnarray}
 f_{B_s} \hat{B}_{B_s}^{1/2}\Big|_{\rm JLQCD} = 0.245 \pm 0.021_{-0.002}^{+0.003}
~\text{GeV},
\nonumber\\
  f_{B_s} \hat{B}_{B_s}^{1/2}\Big|_{\rm (HP+JL)QCD} = 0.295 \pm 0.036 ~ \text{GeV},
\label{hadronic}
 \end{eqnarray}
taken by the lattice calculations \cite{lattice QCD}, we obtain
 \begin{eqnarray}
  \Delta m_s^{\rm SM}\Big|_{\rm JLQCD} = 15.57_{-2.60}^{+2.82} ~ \text{ps}^{-1},
  \nonumber\\
  \Delta m_s^{\rm SM}\Big|_{\rm (HP+JL)QCD} = 22.57_{-5.22}^{+5.88} ~ \text{ps}^{-1},
  \label{eq:dms_SM}
 \end{eqnarray}
respectively.

We will perform the numerical analysis using combined lattice calculations,
${\rm (HP+JL)QCD}$, for CDF experimental result shown in (1),
unless we state otherwise.
Now we investigate the effects of the leptophobic $Z^\prime$ gauge boson.
In the leptophobic $Z^\prime$ model, we have two parameters,
the mass of $Z^{\prime}$ boson and new FCNC coupling, $U_{sb}^{Z^\prime}$.
Since the D{\O} experiment
excludes the mass range $365~{\rm GeV} \leq M_{Z^\prime} \leq 615~{\rm GeV}$
\cite{Abbott:1997dr},
we take $M_{Z^\prime}$ larger than $700$ GeV,
which is also consistent with the mass bound of the conventional
non-leptophobic $Z^\prime$ model.
The coupling $U_{sb}^{Z^\prime}$ has in general CP violating complex phase,
which we denote as $\phi_{sb}^{Z^\prime}$.


The $Z^\prime$-exchanging $\Delta B = \Delta S = 2$ tree diagram
contributes to the $B_s^0-\overline{B}_s^0$ mixing.
The mass difference $\Delta m_s$ of the mixing parameters then read
\begin{eqnarray}
\Delta m_s
= \Delta m_s^{\rm SM}
   \left|1 + R ~e^{2i \phi_{sb}^{Z^\prime}} \right|,
\end{eqnarray}
\begin{eqnarray}
R \equiv  \frac{2\sqrt{2} \pi^2}
     {G_F M_W^2 \left( V_{tb} V{_{ts}^\ast} \right)^2  S_0(x_t)}
     \frac{M_Z^2}{M_{Z^\prime}^2}
     \left|U_{sb}^{Z^\prime}\right|^2
     = 1.62 \times 10^3
       \left(\frac{700 ~\text{GeV}}{M_{Z^\prime}}\right)^2
       \left|U_{sb}^{Z^\prime}\right|^2.
\end{eqnarray}

\begin{figure}
\begin{tabular}{cc}
~~~\psfig{file=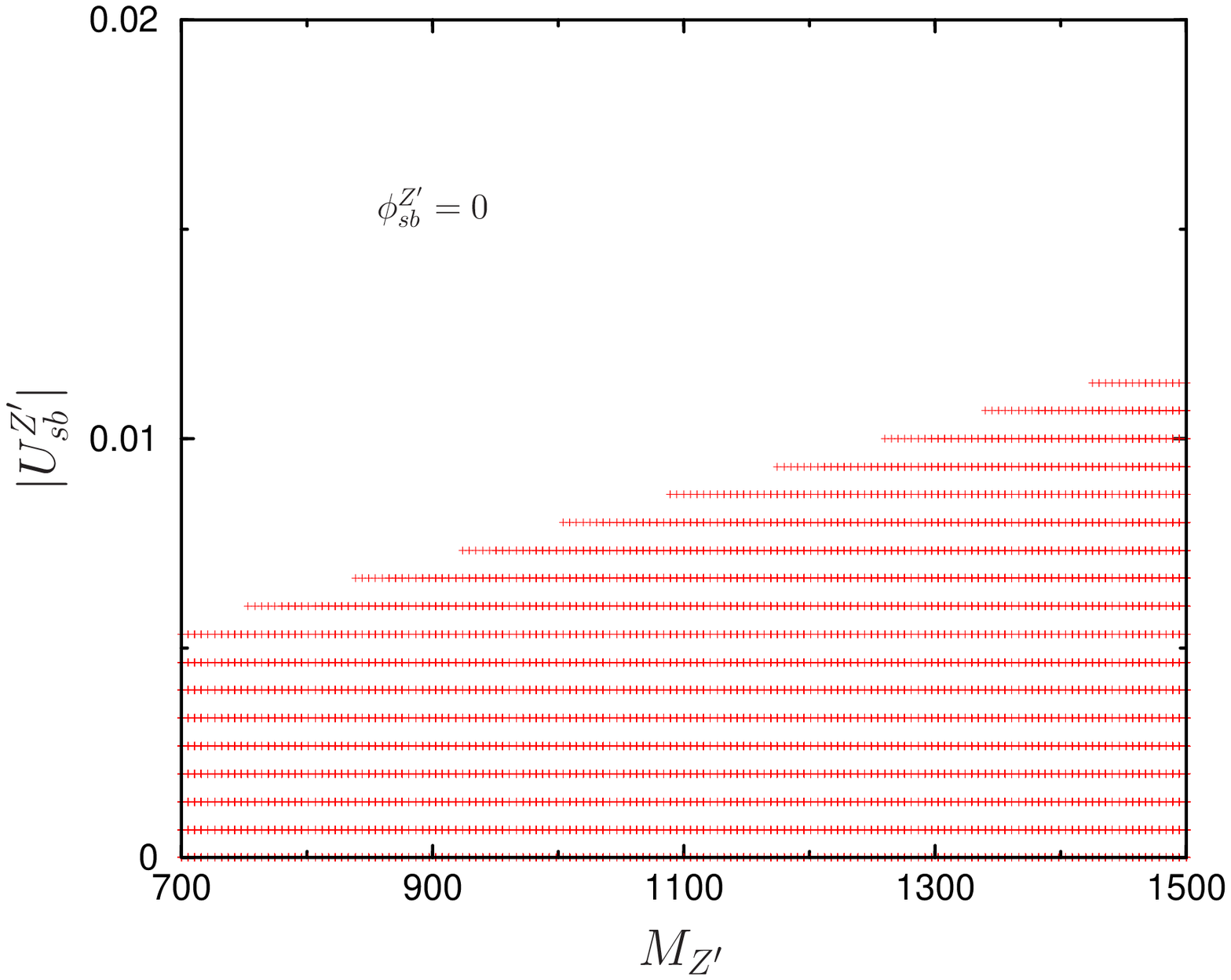,width=7cm}~~~&
~~~\psfig{file=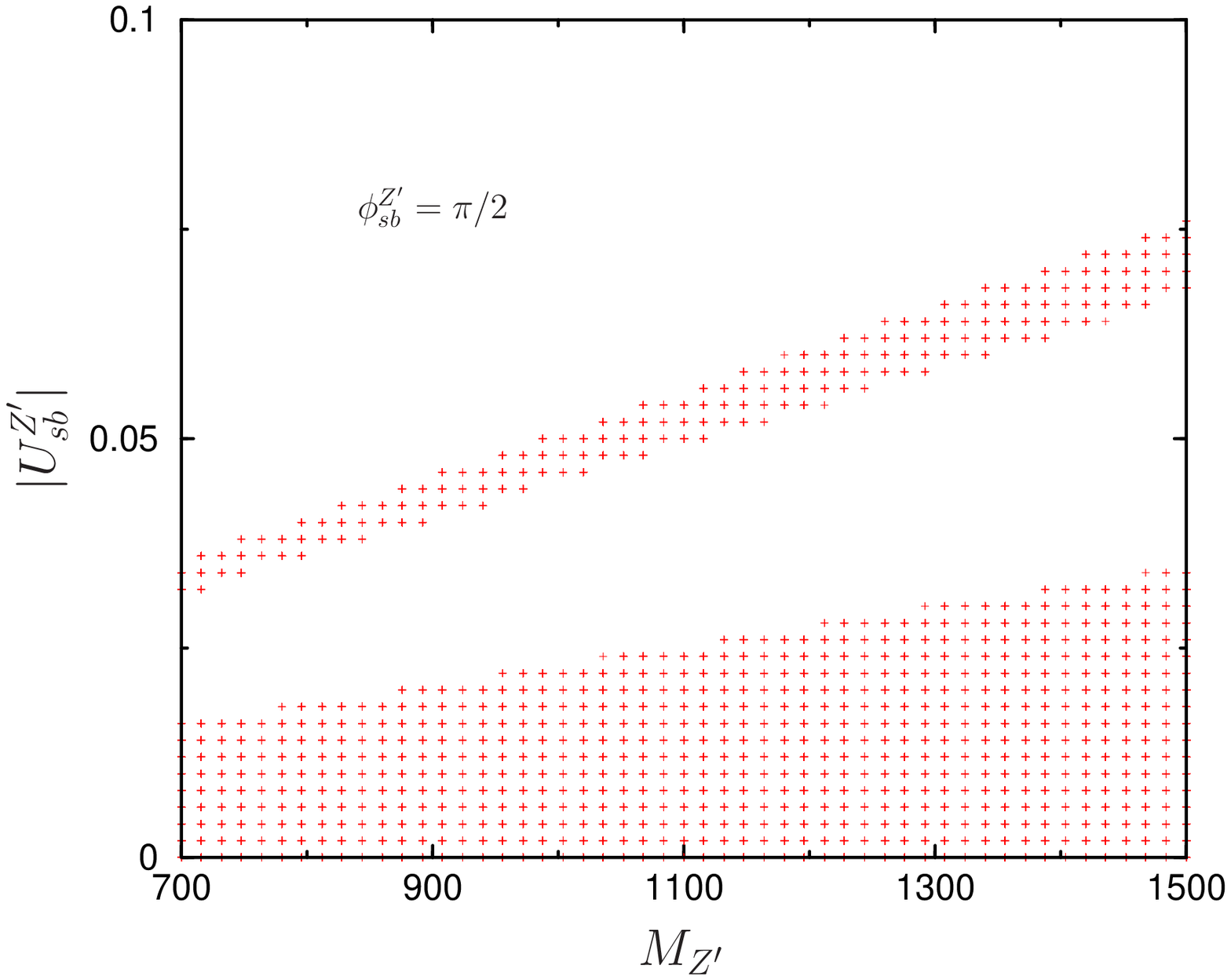,width=7cm}~~~\\[-2.0ex]
\textbf{(a)}&\textbf{(b)}
\end{tabular}
\vspace*{8pt}
\caption{ \label{fig1}
The allowed region in ($M_{Z^\prime}$,$|U_{sb}^{Z^\prime}|$) plane
for (a)~$\phi_{sb}^{Z^\prime}=0$ and
(b)~$\phi_{sb}^{Z^\prime}=\pi/2$~.
}
\end{figure}

In Figs.~\ref{fig1}, we show the allowed region in
($M_{Z^\prime}$,$|U_{sb}^{Z^\prime}|$) plane for vanishing (a) and
maximal (b) phase. We obtain \begin{equation} |U_{sb}^{Z^\prime}|
\leq 0.0055 \qquad
                    \text{for} ~M_{Z^\prime} = 700 ~ \text{GeV},
\end{equation}
for $\phi_{sb}^{Z^\prime}=0$.
This bound is about two orders of magnitude stronger than
the one previously obtained from exclusive semileptonic
$B \to M \nu \bar{\nu}$ decays, $|U_{sb}^{Z^\prime}| \leq 0.29$
\cite{Jeon:2006nq}.
This demonstrates the importance of the measurement of
$B_s^0-\overline{B}_s^0$ mixing in constraining NP in the flavor sector.
Since for $\phi_{sb}^{Z^\prime}=0$ the $Z^\prime$ contribution
is constructive (the same sign) with the SM, the constraint
is very strong.
This severe constraint can be alleviated significantly by allowing the phase
$\phi_{sb}^{Z^\prime}$ to be non-zero.
For maximal phase $\phi_{sb}^{Z^\prime}=\pi/2$,
the $Z^\prime$ contribution is destructive with the SM one and
relatively large value of $|U_{sb}^{Z^\prime}|$
is allowed, as can be seen from Fig.~\ref{fig1}(b),
\begin{equation}
|U_{sb}^{Z^\prime}| \leq 0.036 \qquad
                    \text{for} ~M_{Z^\prime} = 700 ~ \text{GeV},
\end{equation}
irrespective of its phase $\phi_{sb}^{Z^\prime}$ value.
The blank region between the two allowed
regions is excluded by the lower value of $\Delta m_s^{\rm exp}$.
This non-vanishing phase can also contribute to other CP violating processes
such as CP-asymmetries in non-leptonic $B$-meson decays with $b\to s$
transition such as $B \to \pi K$ or $B \to \phi K_S$ decays~\cite{Barger:2004hn,BJK}.

\begin{figure}
\begin{tabular}{cc}
~~~\psfig{file=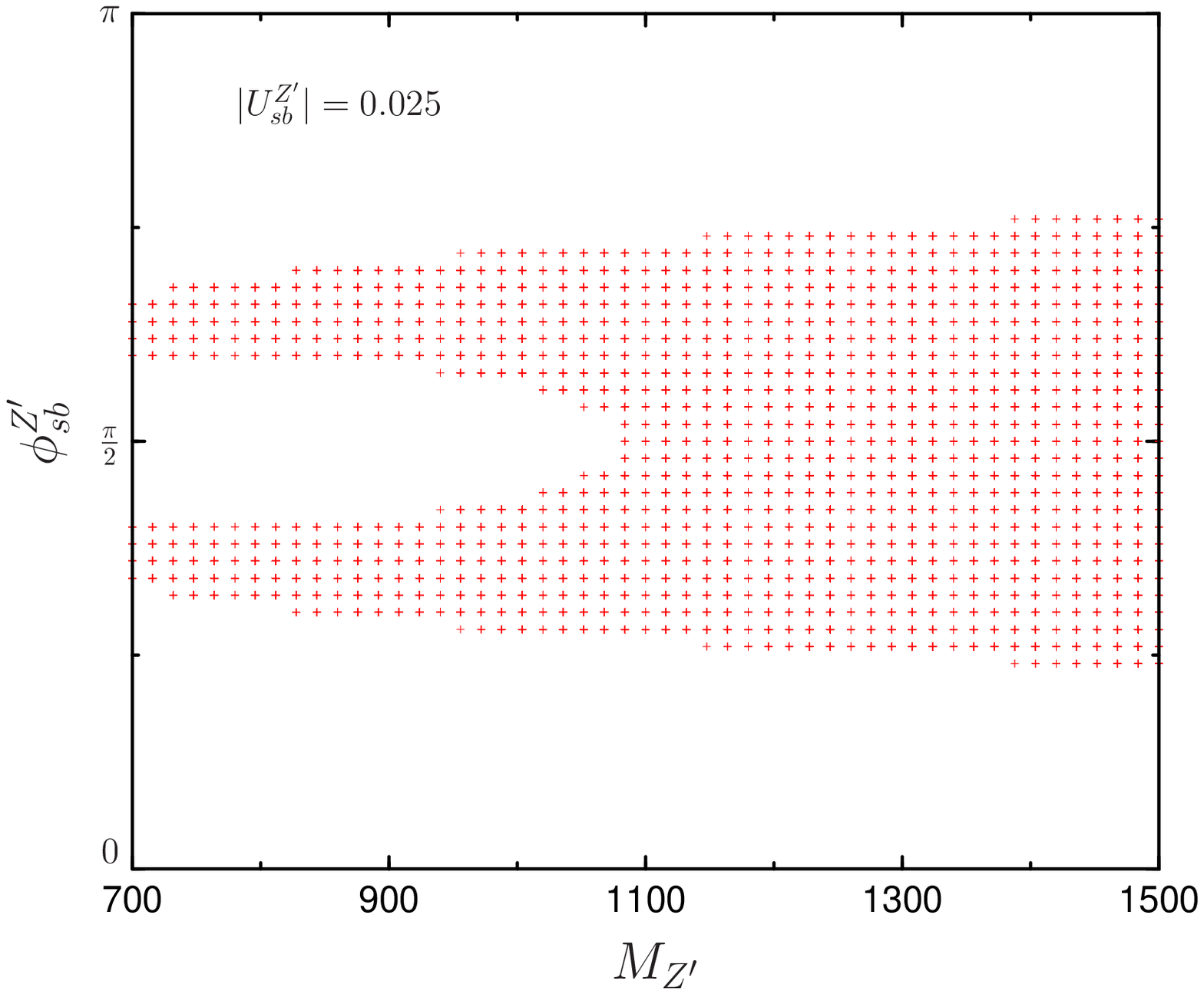,width=7cm}~~~&
~~~\psfig{file=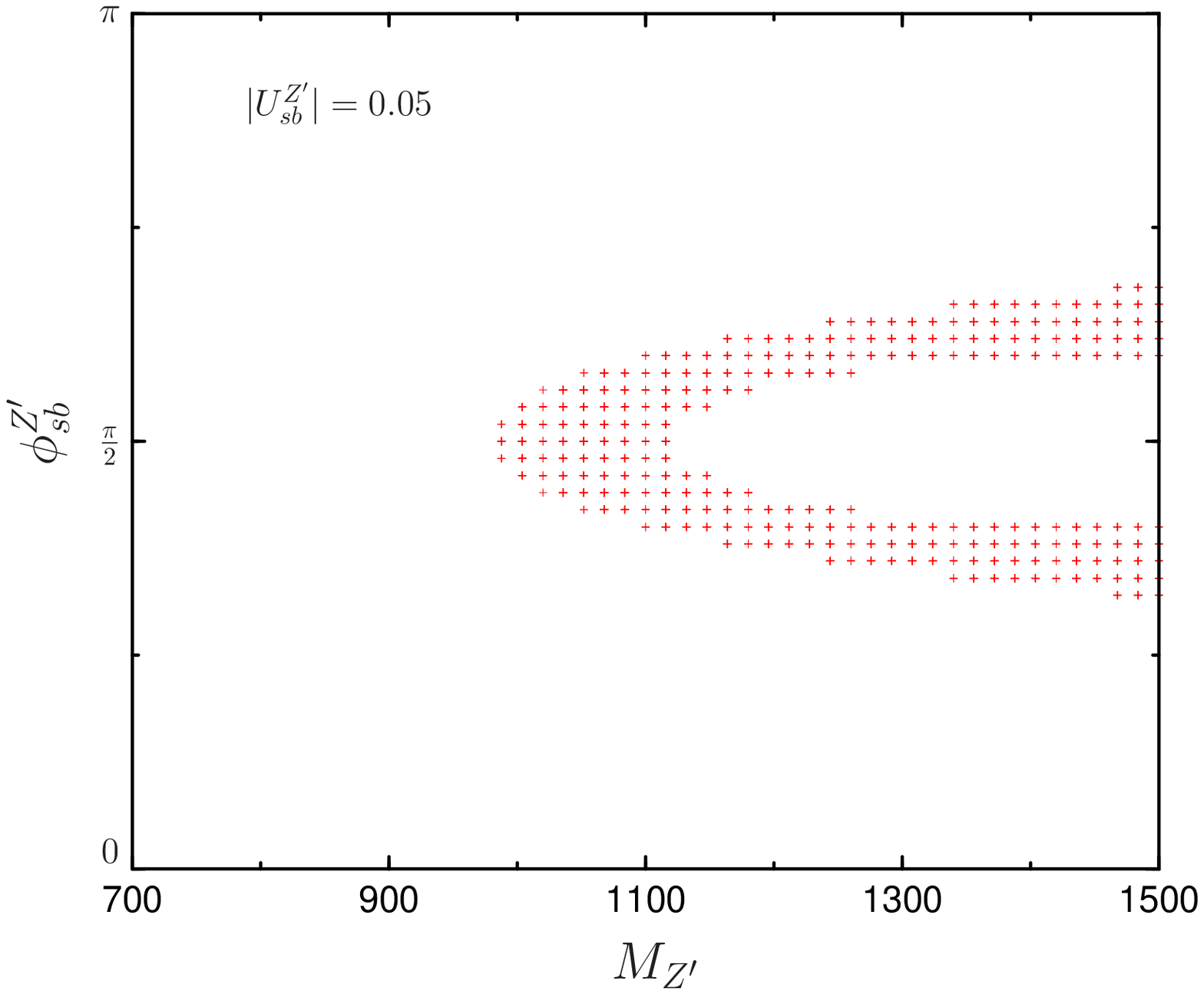,width=7cm}~~~\\[-2.0ex]
\textbf{(a)}&\textbf{(b)}
\end{tabular}
\vspace*{8pt}
\caption{ \label{fig2}
The allowed region in ($M_{Z^\prime}$,$\phi_{sb}^{Z^\prime}$) plane
for (a)~$|U_{sb}^{Z^\prime}|=0.025$ and
(b)~$|U_{sb}^{Z^\prime}|=0.05$~.
}
\end{figure}

In Figs.~\ref{fig2}, we present our predictions for the allowed
$\phi_{sb}^{Z^\prime}$ in the leptophobic $Z^\prime$ model as a
function of $M_{Z^\prime}$ for
(a)~$|U_{sb}^{Z^\prime}|=0.025$ and
(b)~$|U_{sb}^{Z^\prime}|=0.05$~.
For the choice of these relatively large couplings the allowed region
appears only near maximal CP violating phase reflecting again
the destructive interference for these values.
We note that for coupling $|U_{sb}^{Z^\prime}| \gtrsim 0.05$
the $Z^\prime$ mass is larger than 1 TeV for which it
would be difficult to produce it even at LHC.
We also note that for such a large coupling $|U_{sb}^{Z^\prime}| = 0.05$
only very limited parameter space is allowed, such as
\begin{equation}
M_{Z^\prime} = 980 \sim 1120 ~{\rm GeV} \qquad  {\rm for} \qquad
\phi_{sb}^{Z^\prime} \sim \pi/2~.
\end{equation}

\begin{figure}
\begin{tabular}{cc}
~~~\psfig{file=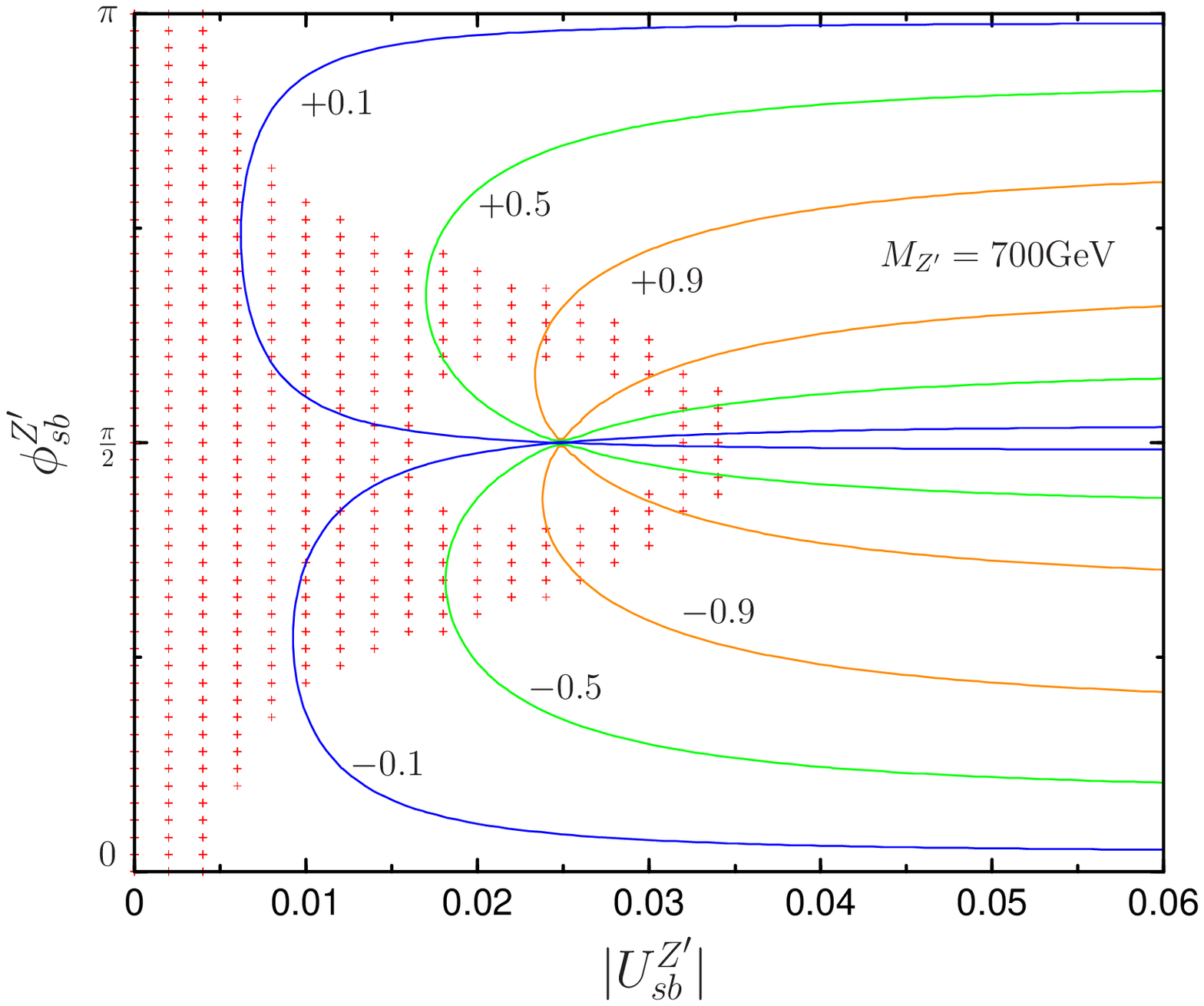,width=7cm}~~~&
~~~\psfig{file=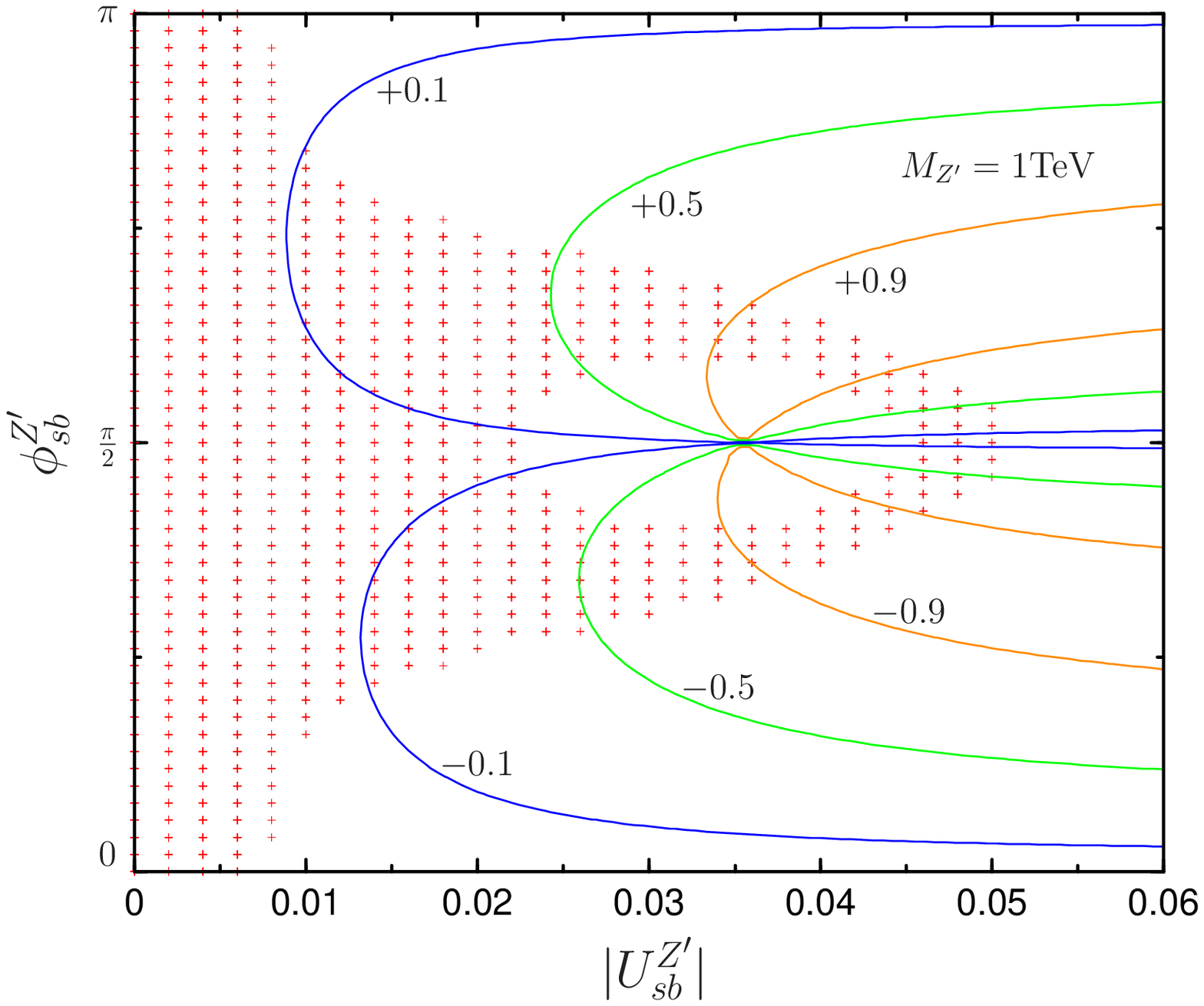,width=7cm}~~~\\[-2.0ex]
\textbf{(a)}&\textbf{(b)}
\end{tabular}
\vspace*{8pt}
\caption{ \label{fig3}
The allowed region in ($|U_{sb}^{Z^\prime}|$,$\phi_{sb}^{Z^\prime}$) plane
for (a)~$M_{Z^\prime}=700$ GeV and
(b)~$M_{Z^\prime}=1$ TeV~. We used (HP+JL)QCD result in (\ref{hadronic}) for the
hadronic parameter. Constant contour lines for the time dependent
CP asymmetry $S_{\psi\phi}$ in $B_s \to J/\psi~\phi$ are also shown.
}
\end{figure}

\begin{figure}
\begin{tabular}{cc}
~~~\psfig{file=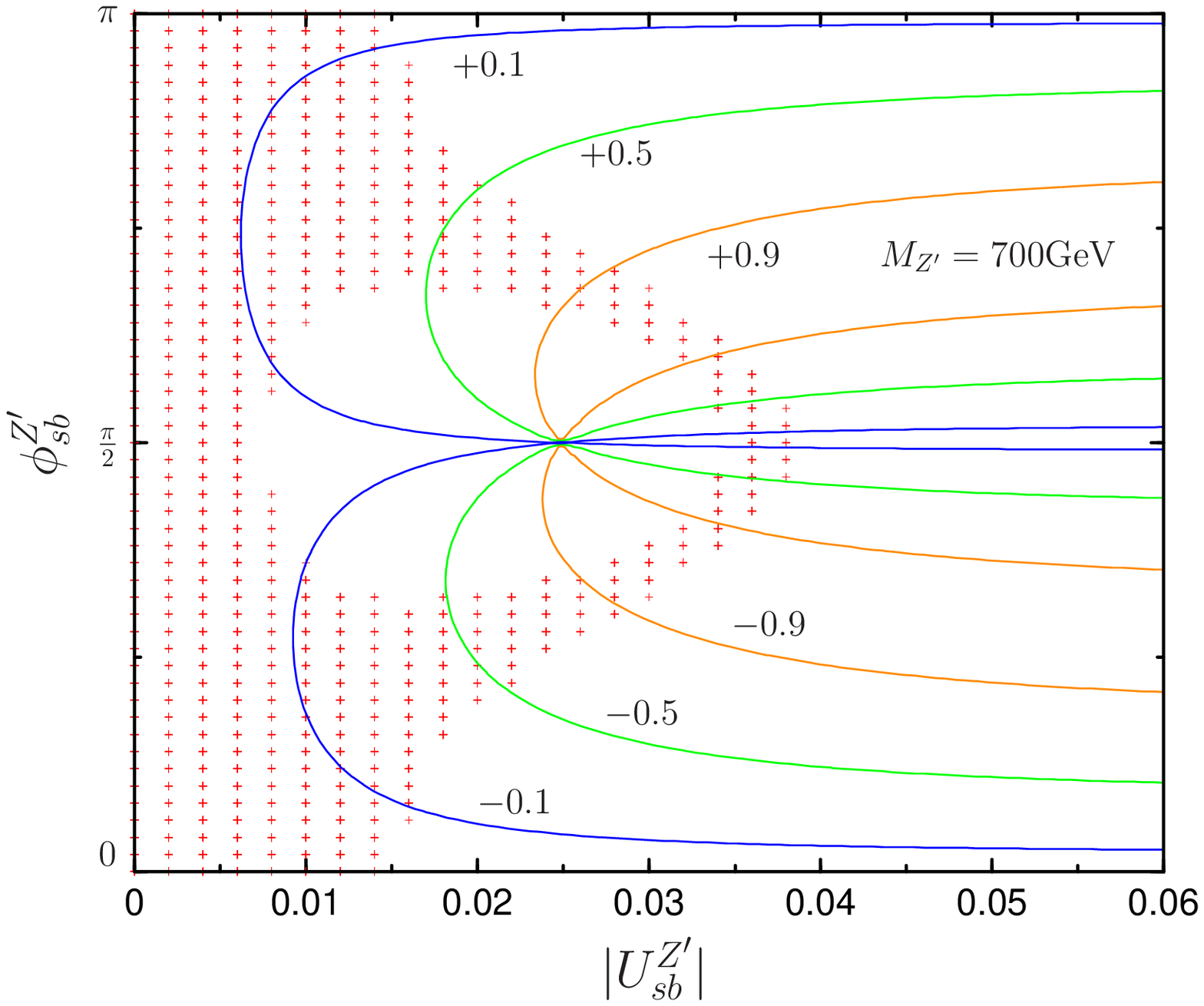,width=7cm}~~~&
~~~\psfig{file=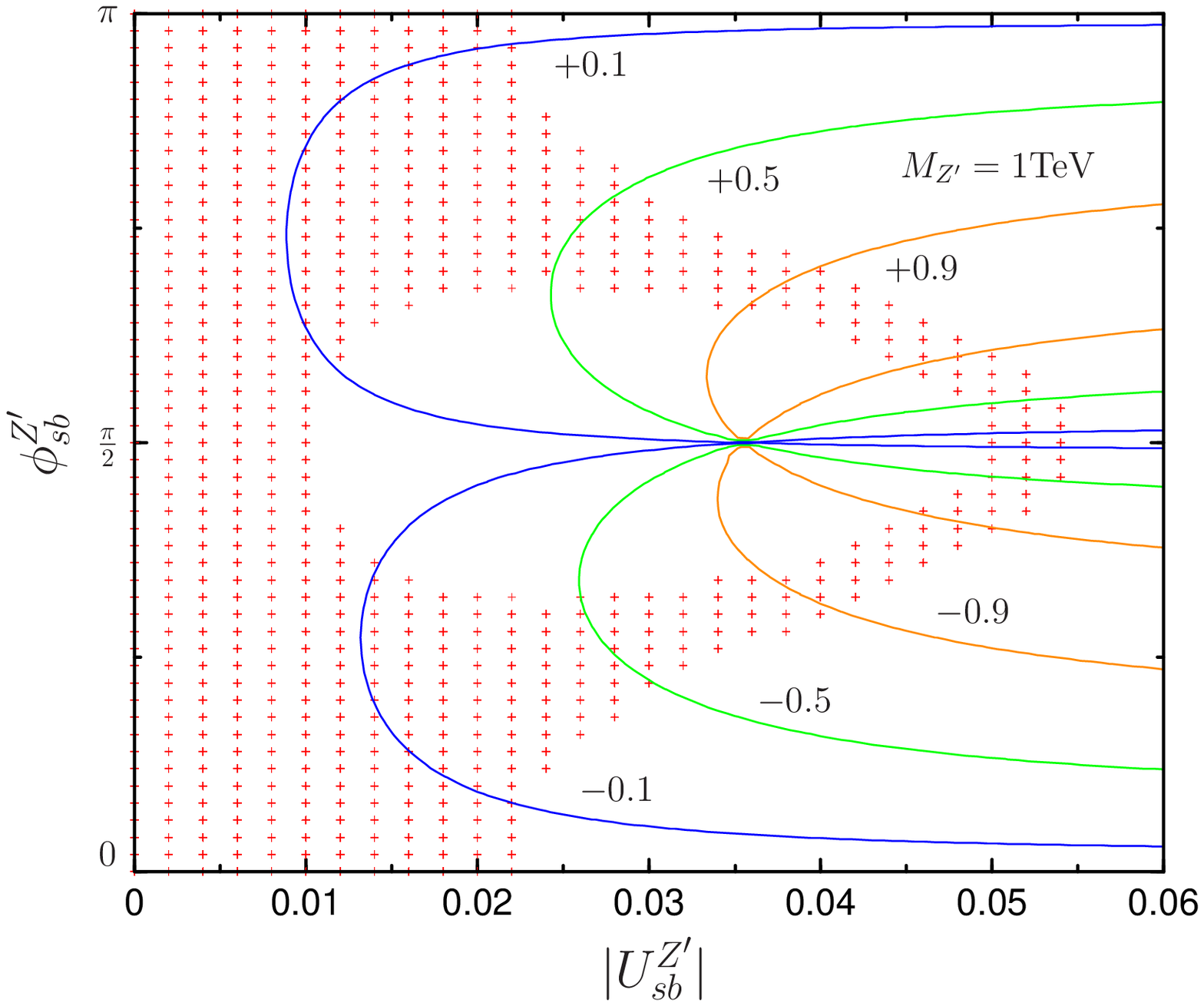,width=7cm}~~~\\[-2.0ex]
\textbf{(a)}&\textbf{(b)}
\end{tabular}
\vspace*{8pt}
\caption{ \label{fig4}
The allowed region in ($|U_{sb}^{Z^\prime}|$,$\phi_{sb}^{Z^\prime}$) plane
for (a)~$M_{Z^\prime}=700$ GeV and
(b)~$M_{Z^\prime}=1$ TeV~. We used JLQCD result in (\ref{hadronic}) for the
hadronic parameter. Constant contour lines for the time dependent
CP asymmetry $S_{\psi\phi}$ in $B_s \to J/\psi~\phi$ are also shown.
}
\end{figure}

In Figs.~\ref{fig3}, the allowed region in
($|U_{sb}^{Z^\prime}|$,$\phi_{sb}^{Z^\prime}$) plane is shown.
The holes again appear because they predict too small $\Delta m_s$.
For a given $M_{Z^\prime}$ we can see that large CP violating phase can enhance
the allowed coupling $|U_{sb}^{Z^\prime}|$ up to almost factor 10.
This shows the importance of the role played by CP violating phase
even in CP conserving observable such as  $\Delta m_s$.
As can be seen from Fig. 3(b),
irrespective of its phase $\phi_{sb}^{Z^\prime}$ value
\begin{equation}
|U_{sb}^{Z^\prime}| \leq 0.051 \qquad
                    \text{for} ~M_{Z^\prime} = 1 ~ \text{TeV}.
\end{equation}

The CP violating phase in $B_s^0 - \overline{B}_s^0$ mixing amplitude
can be measured at LHC in near future through the
time-dependent CP asymmetry in $B_s \to J/\psi~\phi$ decay
\begin{equation}
 \frac{ \Gamma \left(\overline{B}_s^0(t) \to J/\psi~ \phi \right)
       -\Gamma \left(B_s^0(t) \to J/\psi~ \phi \right)}
      { \Gamma \left(\overline{B}_s^0(t) \to J/\psi~ \phi \right)
       +\Gamma \left(B_s^0(t) \to J/\psi~ \phi \right)}
 \equiv S_{\psi\phi} \sin \left(\Delta m_s t\right).
\end{equation}
We note that although the final states are not CP-eigenstates, the time-dependent
analysis of the $B_s^0 \to J/\psi~ \phi$ angular distribution allows a clean
extraction of $S_{\psi\phi}$~\cite{angular}.
In the SM, $S_{\psi\phi}$ is predicted to be very small,
$S_{\psi\phi}^{\rm SM} =-\sin 2\beta_s =0.038 \pm 0.003$
$\left(\beta_s \equiv \arg \left[(V_{ts}^* V_{tb}) / (V_{cs}^*
V_{cb})\right]\right)$. If NP has an additional CP violating phase
$\phi_{sb}^{Z^\prime}$, however, the experimental value of
\begin{equation}
 S_{\psi\phi} = -\sin \left[ 2 \beta_s +
                            \arg \left(1 + R ~e^{2i \phi_{sb}^{Z^\prime}} \right)
                     \right]
\end{equation}
would be significantly different from the SM prediction.
Constant contour lines for $S_{\psi\phi}$ are also shown in Figs.~\ref{fig3}-\ref{fig4}.
We can see that even with the strong constraint from the present $\Delta m_s$ observation,
large $S_{\psi\phi}$ are still allowed.

For comparison, we also show plots similar to Figs.~\ref{fig3} using
other hadronic parameter value ($i.e.$ JLQCD only of Eqs. (10,11))
in Figs.~\ref{fig4}. Since the central value of SM prediction (the
first line in (\ref{eq:dms_SM})) is lower than the CDF central
value, the hole region has been more carved away than in
Figs.~\ref{fig3}. However, we can see that the overall feature is
the same as Figs.~\ref{fig3}.


In this letter we considered leptophobic $Z^\prime$ scenario with
flavor changing neutral current couplings at tree-level. This
scenario can appear in many GUTs such as string-inspired flipped
SU(5)$\times$U(1) models or nonstandard embedding in $E_6$ GUTs
with kinetic mixing.
Since leptophobic $Z^\prime$ does not couple to leptons, the
popular processes for NP searches like $b \to s \ell^+ \ell^-$ or $B_s \to \mu^+ \mu^-$
are not affected by this model.
We showed that the recently measured mass difference $\Delta m_s$
of $B_s^0-\overline{B}_s^0$ system can constrain this kind of models
very efficiently.
The obtained bound on the coupling becomes about two orders
of magnitude stronger than the best known bound.
We also pointed out that the constraint is very sensitive to
the CP violating phase.
Scanning all the possible region of the phase, we obtained
$|U_{sb}^{Z^\prime}| \leq 0.036$ for  $M_{Z^\prime} = 700$ GeV, and
$|U_{sb}^{Z^\prime}| \leq 0.051$ for  $M_{Z^\prime} = 1$ TeV.

\newpage

\centerline{\bf ACKNOWLEDGMENTS}
\noindent We thank Sechul Oh for careful reading of the manuscript and his
valuable comments.
The work of CSK was supported
in part by  CHEP-SRC Program,
in part by the Korea Research Foundation Grant funded by the Korean Government
(MOEHRD) No. KRF-2005-070-C00030.
The work of SB was supported
by the Korea Research Foundation Grant funded by the Korean Government
(MOEHRD) No. KRF-2005-070-C00030.
JHJ was supported by BK21 Program of Korean Government.
\\



\begin{thebibliography}{99}

\bibitem{B2piK}
  A.~J.~Buras, R.~Fleischer, S.~Recksiegel and F.~Schwab,
  Nucl.\ Phys.\ B {\bf 697}, 133 (2004)
  [arXiv:hep-ph/0402112];
  T.~Yoshikawa,
  Phys.\ Rev.\ D {\bf 68}, 054023 (2003)
  [arXiv:hep-ph/0306147];
  V.~Barger, C.~W.~Chiang, P.~Langacker and H.~S.~Lee,
  Phys.\ Lett.\ B {\bf 598}, 218 (2004)
  [arXiv:hep-ph/0406126].
  S.~Nandi and A.~Kundu,
  arXiv:hep-ph/0407061;
    S.~Mishima and T.~Yoshikawa,
  Phys.\ Rev.\ D {\bf 70}, 094024 (2004)
  [arXiv:hep-ph/0408090];
  Y.~Y.~Charng and H.~n.~Li,
  Phys.\ Rev.\ D {\bf 71}, 014036 (2005)
  [arXiv:hep-ph/0410005];
  X.~G.~He and B.~H.~J.~McKellar,
  arXiv:hep-ph/0410098;
  S.~Baek, P.~Hamel, D.~London, A.~Datta and D.~A.~Suprun,
  Phys.\ Rev.\ D {\bf 71}, 057502 (2005)
  [arXiv:hep-ph/0412086];
  Y.~L.~Wu and Y.~F.~Zhou,
  Phys.\ Rev.\ D {\bf 72}, 034037 (2005)
  [arXiv:hep-ph/0503077];
  M.~Gronau and J.~L.~Rosner,
  Phys.\ Rev.\ D {\bf 71}, 074019 (2005)
  [arXiv:hep-ph/0503131];
  C.~S.~Kim, S.~Oh and C.~Yu,
  Phys.\ Rev.\ D {\bf 72}, 074005 (2005)
  [arXiv:hep-ph/0505060];
  S.~Khalil,
  Phys.\ Rev.\ D {\bf 72}, 035007 (2005)
  [arXiv:hep-ph/0505151];
  H.~n.~Li, S.~Mishima and A.~I.~Sanda,
  Phys.\ Rev.\ D {\bf 72}, 114005 (2005)
  [arXiv:hep-ph/0508041];
  R.~Arnowitt, B.~Dutta, B.~Hu and S.~Oh,
  Phys.\ Lett.\ B {\bf 633}, 748 (2006)
  [arXiv:hep-ph/0509233];
  C.~W.~Bauer, I.~Z.~Rothstein and I.~W.~Stewart,
  arXiv:hep-ph/0510241;
  W.~S.~Hou, M.~Nagashima, G.~Raz and A.~Soddu,
  arXiv:hep-ph/0603097;
  S.~Baek,
  arXiv:hep-ph/0605094.
\bibitem{B2roK}
  C.~Dariescu, M.~A.~Dariescu, N.~G.~Deshpande and D.~K.~Ghosh,
  Phys.\ Rev.\ D {\bf 69}, 112003 (2004)
  [arXiv:hep-ph/0308305].
  E.~Alvarez, L.~N.~Epele, D.~G.~Dumm and A.~Szynkman,
  Phys.\ Rev.\ D {\bf 70}, 115014 (2004)
  [arXiv:hep-ph/0410096].
  Y.~D.~Yang, R.~M.~Wang and G.~R.~Lu,
  Phys.\ Rev.\ D {\bf 72}, 015009 (2005)
  [arXiv:hep-ph/0411211].
  P.~K.~Das and K.~C.~Yang,
  Phys.\ Rev.\ D {\bf 71}, 094002 (2005)
  [arXiv:hep-ph/0412313].
  C.~S.~Kim and Y.~D.~Yang,
  arXiv:hep-ph/0412364.
  S.~Baek, A.~Datta, P.~Hamel, O.~F.~Hernandez and D.~London,
  Phys.\ Rev.\ D {\bf 72}, 094008 (2005)
  [arXiv:hep-ph/0508149].
  C.~S.~Huang, P.~Ko, X.~H.~Wu and Y.~D.~Yang,
  Phys.\ Rev.\ D {\bf 73}, 034026 (2006)
  [arXiv:hep-ph/0511129].
\bibitem{B2phiK}
  B.~Dutta, C.~S.~Kim and S.~Oh,
  Phys.\ Rev.\ Lett.\  {\bf 90}, 011801 (2003)
  [arXiv:hep-ph/0208226].
  G.~L.~Kane, P.~Ko, H.~b.~Wang, C.~Kolda, J.~h.~Park and L.~T.~Wang,
  Phys.\ Rev.\ D {\bf 70}, 035015 (2004)
  [arXiv:hep-ph/0212092];
  S.~Baek,
  Phys.\ Rev.\ D {\bf 67}, 096004 (2003)
  [arXiv:hep-ph/0301269];
  B.~Dutta, C.~S.~Kim, S.~Oh and G.~h.~Zhu,
  Phys.\ Lett.\ B {\bf 601}, 144 (2004)
  [arXiv:hep-ph/0312389];
  M.~Endo, M.~Kakizaki and M.~Yamaguchi,
  Phys.\ Lett.\ B {\bf 594}, 205 (2004)
  [arXiv:hep-ph/0403260];
  J.~F.~Cheng, C.~S.~Huang and X.~H.~Wu,
  Nucl.\ Phys.\ B {\bf 701}, 54 (2004)
  [arXiv:hep-ph/0404055];
  S.~Khalil,
  Mod.\ Phys.\ Lett.\ A {\bf 19}, 2745 (2004)
  [Afr.\ J.\ Math.\ Phys.\  {\bf 1}, 101 (2004)]
  [arXiv:hep-ph/0411151].
\bibitem{Bs2mumu}
  C.~S.~Huang and Q.~S.~Yan,
  Phys.\ Lett.\ B {\bf 442}, 209 (1998)
  [arXiv:hep-ph/9803366];
  K.~S.~Babu and C.~F.~Kolda,
  Phys.\ Rev.\ Lett.\  {\bf 84}, 228 (2000)
  [arXiv:hep-ph/9909476];
  P.~H.~Chankowski and L.~Slawianowska,
  Phys.\ Rev.\ D {\bf 63}, 054012 (2001)
  [arXiv:hep-ph/0008046];
  G.~Isidori and A.~Retico,
  JHEP {\bf 0111}, 001 (2001)
  [arXiv:hep-ph/0110121];
  C.~Bobeth, T.~Ewerth, F.~Kruger and J.~Urban,
  Phys.\ Rev.\ D {\bf 64}, 074014 (2001)
  [arXiv:hep-ph/0104284];
  S.~Baek, P.~Ko and W.~Y.~Song,
  Phys.\ Rev.\ Lett.\  {\bf 89}, 271801 (2002)
  [arXiv:hep-ph/0205259];
  JHEP {\bf 0303}, 054 (2003)
  [arXiv:hep-ph/0208112];
   S.~Baek,
  Phys.\ Lett.\ B {\bf 595}, 461 (2004)
  [arXiv:hep-ph/0406007].
\bibitem{D0}
  V.~Abazov  [D0 Collaboration],
  arXiv:hep-ex/0603029.
\bibitem{CDF}
  G.~Gomez-Ceballos [CDF Collaboration], Talk at FPCP 2006,
  http://fpcp2006.triumf.ca/agenda.php.
\bibitem{model_indep1}
  M.~Blanke, A.~J.~Buras, D.~Guadagnoli and C.~Tarantino,
  arXiv:hep-ph/0604057;
  Z.~Ligeti, M.~Papucci and G.~Perez,
  arXiv:hep-ph/0604112;
  P.~Ball and R.~Fleischer,
  arXiv:hep-ph/0604249;
  Y.~Grossman, Y.~Nir and G.~Raz,
  arXiv:hep-ph/0605028;
  A.~Datta,
  arXiv:hep-ph/0605039.
\bibitem{zprime}
  K.~Cheung, C.~W.~Chiang, N.~G.~Deshpande and J.~Jiang,
  arXiv:hep-ph/0604223;
  X.~G.~He and G.~Valencia,
  arXiv:hep-ph/0605202.

\bibitem{MSSM}
  M.~Ciuchini and L.~Silvestrini,
  arXiv:hep-ph/0603114;
  M.~Endo and S.~Mishima,
  arXiv:hep-ph/0603251.
  J.~Foster, K.~i.~Okumura and L.~Roszkowski,
  arXiv:hep-ph/0604121;
    G.~Isidori and P.~Paradisi,
  arXiv:hep-ph/0605012.
  S.~Khalil,
  arXiv:hep-ph/0605021;
  S.~Baek,
  arXiv:hep-ph/0605182;
  R.~Arnowitt, B.~Dutta, B.~Hu and S.~Oh,
  arXiv:hep-ph/0606130.

\bibitem{RS-kim}
S.~Chang, C.~S.~Kim and J.~Song,
  arXiv:hep-ph/0607313.

\bibitem{Lopez:1996ta}
  J.~L.~Lopez and D.~V.~Nanopoulos,
  Phys.\ Rev.\ D {\bf 55}, 397 (1997)
  [arXiv:hep-ph/9605359].

\bibitem{Rizzo:1998ut}
T.~G.~Rizzo,
  Phys.\ Rev.\ D {\bf 59}, 015020 (1999)
  [arXiv:hep-ph/9806397].

\bibitem{Babu:1996vt}
  K.~S.~Babu, C.~F.~Kolda and J.~March-Russell,
  Phys.\ Rev.\ D {\bf 54}, 4635 (1996)
  [arXiv:hep-ph/9603212];
  {\it ibid.} {\bf 57}, 6788 (1998)
  [arXiv:hep-ph/9710441].

\bibitem{Leroux:2001fx}
  K.~Leroux and D.~London,
  Phys.\ Lett.\ B {\bf 526}, 97 (2002)
  [arXiv:hep-ph/0111246].

\bibitem{Jeon:2006nq}
  J.~H.~Jeon, C.~S.~Kim, J.~Lee and C.~Yu,
  Phys.\ Lett.\ B {\bf 636}, 270 (2006)
  [arXiv:hep-ph/0602156].

\bibitem{Charles:2004jd}
  J.~Charles {\it et al.}  [CKMfitter Group],
  Eur.\ Phys.\ J.\ C {\bf 41}, 1 (2005)
  [arXiv:hep-ph/0406184].


\bibitem{lattice QCD}
  S.~Aoki {\it et al.}  [JLQCD Collaboration],
  Phys.\ Rev.\ Lett.\  {\bf 91}, 212001 (2003)
  [arXiv:hep-ph/0307039];
  A.~Gray {\it et al.}  [HPQCD Collaboration],
  Phys.\ Rev.\ Lett.\  {\bf 95}, 212001 (2005)
  [arXiv:hep-lat/0507015];
  M.~Okamoto,
  PoS {\bf LAT2005}, 013 (2006)
  [arXiv:hep-lat/0510113];
  P.~Ball and R.~Fleischer,
  arXiv:hep-ph/0604249.


\bibitem{Abbott:1997dr}
  B.~Abbott {\it et al.}  [D0 Collaboration],
FERMILAB-CONF-97-356-E
  {\it Presented at 18th International Symposium on Lepton and Photon Interactions (LP 97), Hamburg, Germany, 28 Jul - 1 Aug 1997, and Presented at
  International Europhysics Conference on High-Energy Physics (HEP 97), Jerusalem, Israel, 19-26 Aug 1997}

\bibitem{Barger:2004hn}
  V.~Barger, C.~W.~Chiang, P.~Langacker and H.~S.~Lee,
  Phys.\ Lett.\ B {\bf 598}, 218 (2004)
  [arXiv:hep-ph/0406126].

\bibitem{BJK}
 S.~Baek, J.~H.~Jeon and C.~S.~Kim, work in preparation.

 \bibitem{angular}
  A.~S.~Dighe, I.~Dunietz and R.~Fleischer,
  Eur.\ Phys.\ J.\ C {\bf 6}, 647 (1999)
  [arXiv:hep-ph/9804253];
  I.~Dunietz, R.~Fleischer and U.~Nierste,
  Phys.\ Rev.\ D {\bf 63}, 114015 (2001)
  [arXiv:hep-ph/0012219].




\end{thebibliography}
\end{document}